\documentclass[twocolumn]{aastex62}
\usepackage{multirow}
\usepackage[normalem]{ulem}
\usepackage{enumitem}

\usepackage{mathtools}

\graphicspath{{./}{figures/}}

\begin{document}

\title{Searching for Young Stellar Objects through SEDs by Machine Learning}

\author{Yi-Lung Chiu${}^\dagger$}
\affil{Institute of Astronomy, National Tsing Hua University, Hsinchu 30013, Taiwan}
\author{Chi-Ting Ho${}^\dagger$}
\affil{Physics Department, National Tsing Hua University, Hsinchu 30013, Taiwan}
\author{Daw-Wei Wang}
\affil{Physics Department, National Tsing Hua University, Hsinchu 30013, Taiwan}
\affil{Physics Division, National Center for Theoretical Sciences, Hsinchu 30013, Taiwan}
\affil{Center for Quantum Technology, National Tsing Hua University, Hsinchu 30013, Taiwan}
\author{Shih-Ping Lai}
\affil{Institute of Astronomy, National Tsing Hua University, Hsinchu 30013, Taiwan}
\affil{Physics Department, National Tsing Hua University, Hsinchu 30013, Taiwan}

\footnote{${}^\dagger$ These authors contributed equally to this work}

%%%%%%%%%%%%%%%%%%%%%%
\begin{abstract}

Accurate measurements of statistical properties, such as the star formation rate and the lifetime of young stellar objects (YSOs) in different stages, is essential for constraining star formation theories. However, it is a difficult task to separate galaxies and YSOs based on spectral energy distributions (SEDs) alone, because they contain both thermal emission from stars and dust around them and no reliable theories can be applied to distinguish them. 
Here we compare different machine learning algorithms and develop the Spectrum Classifier of Astronomical Objects (SCAO), based on Fully Connected Neural Network (FCN), to classify regular stars, galaxies, and YSOs. Superior to previous classifiers, SCAO is solely trained by high quality data labeled in Molecular Cores to Planet-forming Disks (c2d) catalog without a priori theoretical knowledge, and provides excellent results with high precision ($>$96\%) and recall ($>$98\%) for YSOs when only eight bands are included. 
We systematically investigate the effects of observation errors and distance effects, and show that high accuracy performance is still maintained even when using fluxes of only three bands (IRAC 3, IRAC 4, and MIPS 1) in the long wavelengths regime, because the silicate absorption feature is automatically detected by SCAO. Finally, we apply SCAO to \textit{Spitzer} Enhanced Imaging Products (SEIP), the most complete catalog of \textit{Spitzer} observations, and found 129219 YSO candidates. The website from SCAO is available at http://scao.astr.nthu.edu.tw.
\end{abstract}


%%%%%%%%%%%%%%%%%%%%%%%%%%%%%%%%%%%%%%%%%%%%%%%%%%%%%%%%%%%
\section{Introduction} \label{sec:intro}

The timescale of forming a star from the gravitational collapse of molecular clouds is estimated by various theories to be a few 10$^6$-10$^7$ years depending on the stellar mass and the environment, such as whether the clouds are supported by magnetic fields and/or influenced by shocks from a nearby supernova explosion 
\citep{McKee2007, Bergin2007}.
From an observational point of view, since we are unable to follow the formation of a single star, it is essential to observe young stellar objects (YSOs) in different evolutionary stages to constrain theories.
Although there have been many YSOs observed in IRAC, \textit{Spitzer} did make great progress with its high angular resolution and its superior sensitivity compared to IRAS in mid-infrared wavelengths \citep{Werner2004}.
%However, YSOs are mostly embedded in cold molecular clouds, and hence the ability to collect a large amount of YSO data was not achievable until the launch of the \textit{Spitzer} Space Telescope that provides high-sensitivity measurements in mid-infrared wavelengths \citep{Werner2004}.
However, it was soon found that the \textit{Spitzer} Space Telescope is sensitive enough to detect numerous background galaxies with their SEDs similar to the SEDs of YSOs 
\citep{Harvey2007, Hsieh2013}. 
Several methods have been proposed to separate galaxies and YSOs based on their distributions in color-color diagrams and color–magnitude diagrams \citep{Gutermuth2005, Harvey2007, Rebull2010}. 
Here we aim at testing whether machine learning can be a useful tool for identifying the nature of astronomical sources with limited numbers of photometric observations in their SEDs.

Machine learning has been successfully applied in many areas of our daily life, as well as in fundamental research. The popularity has grown significantly in recent years because of rapid developments in computational algorithms, high speed processors, and big data availability from various resources \citep{Russell2014}. These factors also make deep learning possible in artificial neural networks, 
so that hidden patterns may be more efficiently recognized after proper training through big data than by human knowledge/experience.
From this point of view, searching for new YSOs from well-trained models is a reasonable approach compared to traditional theoretical methods, when the amount of observational data is too much to be analyzed by humans alone. 
This method is mostly based on a supervised learning process, where the machine learning model is trained by certain labeled data and then applied to classify/predict other data, which may not be fully understood yet in the previous works.

A few machine learning algorithms related to YSO research have been used, including Support Vector Machine (SVM) \citep{Miettinen2018, deVilliers2013, Marton2016, Marton2019}, Random Forest (RF) \citep{Miettinen2018, Hedges2018, Marton2019}, Neural Network (NN) \citep{Miettinen2018, Marton2019}, Gradient Boosting Machine (GBM) \citep{Miettinen2018}, Decision Tree (DT)\citep{Miettinen2018, Akras2019}, K-Nearest Neighbours (KNN) \citep{Miettinen2018, Marton2019}, 
logistic regression \citep{Miettinen2018}, and naive Bayes classifier \citep{Miettinen2018, Marton2019}, etc. The input data include SEDs \citep{Miettinen2018, Akras2019, Marton2016, Marton2019}, light curves \citep{Hedges2018}, and outflow morphologies \citep{deVilliers2013}. Most of the SED classifications are carried out by using the 2MASS and AllWISE catalogs \citep{Akras2019, Marton2016, Marton2019}, since both of them are all sky surveys with a huge amount of data.

Among these papers, \citet{Marton2016, Marton2019} provide the largest number of YSOs. However, they adopt all previously identified YSOs determined with different criteria, which may not be self-consistent.
In order to obtain more unambiguously confirmed YSO sources to derive the star formation rate in molecular clouds under various conditions \citep{Chiu_in_prep},
it is therefore better to train a machine learning model solely based on a self-consistent theory and verified by the ``ground truth" independently. Here, ground truth is an idea in machine learning indicating the correct answer to a specific question. Furthermore, for a deeper physical understanding of star formation, it is also necessary to identify the most important features for a YSO. Therefore a systematic comparison between different features are critical.

In this paper, we develop a supervised machine learning algorithm, Spectrum Classifier of Astronomical Objects (SCAO) solely trained from SEDs to classify three basic types of astronomical objects: star, galaxy and YSO. The dataset of our SED sources are composed of observational data from  2MASS, UKIDSS, and \textit{Spitzer} (see Section  \ \ref{sec:data}). 
The underlying algorithm of SCAO is built on a FCN, since we find it performs slightly better and more stable than other algorithms after a careful comparison.
%It is well-known that a higher performance machine learning model relies not only on the amount of data, but also strongly relies on the quality of the data. It is usually believed that \textit{Spitzer} can provide a better result than ALLWISE at mid-infrared wavelengths due to the longer exposure time giving a higher signal to noise ratio, although ALLWISE provides a fairly larger number of samples for YSO classification \citep{Marton2019}.
Using data carefully labeled by the previous work (\citet{Evans2007}, which is hereafter mentioned as the c2d method in this paper), SCAO could be used to identify stars and galaxies with very high precision and recall, and to identify YSOs with precision more than 96\% and recall more than 98\% in the test data, fully reproducing previous results without any a priori knowledge.

After successfully reproducing the previous results with SCAO, we compared results obtained by different models to understand the most important features of these SEDs.
For example, we show that the classification can be also well-performed when the distance effect is fully removed by using normalized SEDs. This indicates that the different physical structures of galaxies and YSOs can still be well recognized by the normalized SED, even if their normalized SED profiles are quite similar. A high prediction accuracy for YSOs can also be retrieved by using SEDs comprised of only three bands (IRAC 3, IRAC 4, and MIPS 1), where the extinction effect is almost negligible. This implies that even though these broad-band observations do not cover the most crucial regime for the identification of silicate absorption \citep{Draine2003}, which has a wavelength range $10\mu {\rm m}<\lambda< 20 \mu {\rm m}$, our SCAO could still correctly capture the essential hidden properties from those SEDs nearby to distinguish YSOs from galaxies and stars.

We also compare the predicted results to sources in extragalactic regions, as well as to known YSOs, and find an extremely good agreement with these ground truth. Finally, we find 129219 YSO candidates when using \textit{Spitzer} Enhanced Imaging Products (SEIP), which is the most complete catalog of \textit{Spitzer} observations with $\sim$42M samples.

%%%%%%%%%%%%%%%%%%%%%%%%%%%%%%%%%%%%%%%%%%
\section{Data}
\label{sec:data}

Throughout this paper, we train and test our models to classify infrared sources by using their SEDs. We include eight wavelength bands in total, covering J, H, and K/$K_{s}$ in the near infrared, four IRAC bands and the MIPS 1 band in the mid-infrared (Table \ref{tab:bands}). 
These bands contain emission from both stellar components and dust components which can help to distinguish YSOs from galaxies and/or stars.  
%=================
\begin{figure}[htb!]
\centering
\includegraphics[width=0.48\textwidth]{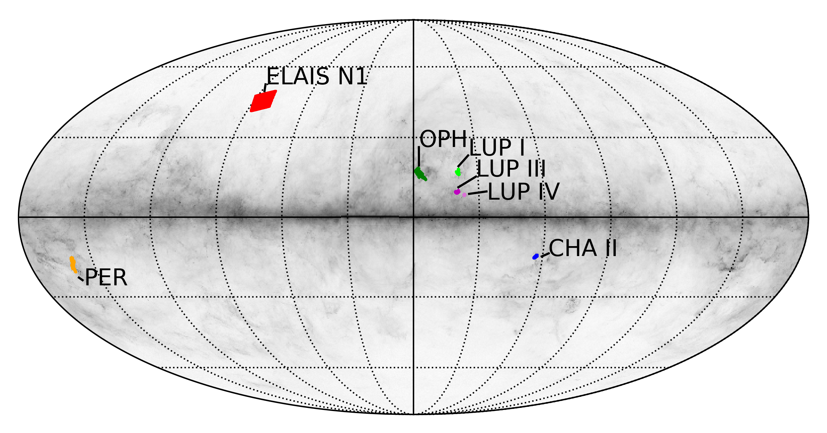}
\caption{Locations of datasets in the all sky map. The background image shown in the galactic coordinate system is dust emission at 545GHz from the Planck Legacy Archive. ElAIS N1 is located off the galactic plane, where almost no YSOs exist. The other regions are near molecular clouds and have more YSOs.
}\label{fig:regions_in_all_sky_map}
\end{figure}
%=================
%=====================Table=======================%
\begin{table}
\centering
\begin{tabular}{c c c c c}
\hline\hline
Observatories& Band systems & $\lambda$ & project\\
 & & ($\mu m$) &\\
\hline
2MASS/UKIRT& UKIRT J & 1.235 & 2MASS/UKIDSS \\
2MASS/UKIRT& UKIRT H & 1.662 & 2MASS/UKIDSS \\
2MASS/UKIRT& UKIRT K & 2.159 & 2MASS/UKIDSS\\
\hline
& IRAC 1 & 3.550 & c2d/SWIRE\\
& IRAC 2 & 4.493 & c2d/SWIRE\\
& IRAC 3 & 5.731 & c2d/SWIRE\\
& IRAC 4 & 7.872 & c2d/SWIRE\\
& MIPS 1 & 24.00 & c2d/SWIRE\\
\hline
\end{tabular}
\caption{\label{tab:bands}The eight bands of the SEDs used in this paper, where $\lambda$ indicates the central wavelength.}
\end{table}
%=================
%%%%%%%%%%%%%%%%%%%%%%%%%%%%%%%%%%%%%%
\subsection{Sources of data }
\label{sec:source_of_data}

The sources we used to test and validate our SCAO models are from two \textit{Spitzer} legacy programs, Molecular Cores to Planet-forming Disks (c2d) and the \textit{Spitzer} Wide-Area Infrared Extragalactic (SWIRE) Survey.
The c2d legacy program, led by \citet{Evans2003}, aimed to study the stellar content of five nearby star-forming molecular clouds, Chamaeleon II, Lupus I, Lupus III, Lupus IV, Ophiuchus, and Perseus (hereafter CHA II, LUP I, LUP III, LUP IV, OPH and PER respectively). 
The SWIRE legacy program, led by \citet{Lonsdale2003}, surveyed several extragalactic regions to study the evolution of dusty star-forming galaxies, evolved stellar populations, and active galactic nuclei; here we only use the ELAIS N1 region since it is the only one in the SWIRE survey used by the c2d method (see Section \ref{sec:actual_labels}).
The locations of these regions are shown in Figure \ref{fig:regions_in_all_sky_map}.
After training SCAO, we apply it to \textit{Spitzer} Enhance Image Products (SEIP) \footnote{\href{https://irsa.ipac.caltech.edu/data/Enhanced/SEIP/docs/seip\_explanatory\_supplement\_v3.pdf}{https://irsa.ipac.caltech.edu/data/Enhanced/ SEIP/docs/seip\_explanatory\_supplement\_v3.pdf}}, which lists all detections with signal to noise ratios ($S/N$) higher than 10 from all available \textit{Spitzer} images.

These programs provide catalogs that contain the observations of  sources in four mid-infrared bands
(3.6$\mu m$, 4.4$\mu m$, 5.7$\mu m$, 8.0$\mu m$) of the IRAC instrument, and the 24$\mu m$ band of the MIPS instrument \citep{Werner2004}.
To extend the SED to the near infrared, 
we also obtain measurements of our sources in J, H, and K/$K_{s}$ bands from the Two Micron All Sky Survey (2MASS) \citep{Skrutskie2006} and the 10th data release of the UKIRT Infrared Deep Sky Survey (UKIDSS)
\footnote{The UKIDSS project is defined in \citet{Lawrence2007}. UKIDSS uses the UKIRT Wide Field Camera (WFCAM; \citet{Casali2007}) and a photometric system described in \citet{Hewett2006}. The pipeline processing and science archive are described in \citet{Irwin_in_prep} and \citet{Hambly2008}.}.
These two surveys use filters with different response functions, central wavelength $\lambda$, and bandwidth $\Delta\lambda$. 
To make the measurements consistent, we convert the 2MASS measurements to the UKIDSS band system using the calibration function described in \citet{Hodgkin2009}.

%%%%%%%%%%%%%%%%%%%%%%
\subsection{Actual labels for our model}
\label{sec:actual_labels}

In order to perform supervised learning to train our model, we adopt source types from \citet{Evans2007} as our actual labels. They define three basic source types for sources with enough number of detections: star, galaxy, and YSO candidate. 
Stars are selected by stellar templates, galaxies are the sources occupying the same regions in several color-color diagrams and color-magnitude diagrams with extragalactic sources, and YSOs are the non-galaxy sources in the same s and color-magnitude diagrams.
Their approaches to distinguish these three types of sources are briefly described as follows: They first selected stars by fitting stellar SED templates. Secondly, they construct a pure galaxy sample by excluding stars from the extragalacitc region, ELAIS N1. 
Thirdly, they construct an unnormalized galaxy probability function based on the distribution of the pure galaxy sample in several color-color diagrams and color-magnitude diagrams (see Fig. 3 and Table 1 in \citet{Harvey2007}); galaxies are selected if the probability is greater than a certain number, then the rest of sources are defined to be YSO candidates.

From the description above, we can see that the sources classified as ``star" by this method contain not only main-sequence stars but also giant stars (i.e. evolved stars).
The sources classified as ``galaxy" by this method should have SEDs similar to extra-galactic sources in ELAIS N1 regions. Finally, the sources classified as ``YSO" by this method should have flux excess at infrared wavelengths and SEDs not similar to those of extra-galactic sources.
Here, YSOs include all evolutionary stages from Class I, Flat, Class II, to Class III, defined by \citet{Greene1994}. However, we note that this YSO list misses pre-main-sequence stars(PMS) that no longer have infrared excess \citep{Evans2009}.

%%%%%%%%%%%%%
\subsection{Data preparation}
\label{sec:data_prep}

We note that catalogs provide upper limits of fluxes for those non-detections, but these upper limits could not provide reliable information for the classification of YSOs. For example, some the upper limits (usually from MIPS 1) could be much higher than the measurements of other bands, which could seriously bias the classification results if one treats them the same as the observed flux. As a result, in this work, we handle the data preparation in a more systematic process, using three different criteria for the SEDs: 1) we require that all four IRAC bands (see Table \ref{tab:bands}) have detections; 2) the MIPS 1 band, the longest wavelength regime, is allowed to have non-detections because we know most stars ($>99\%$) are usually not detectable in this wavelength; 3) the JHK bands, the three bands with the shortest wavelengths, are required to have at least one band detected because YSOs are generally faint in JHK bands and the available number of YSO sources is much less compared to stars.
After applying the data preparation, we obtain 81436 sources in total in our database, where 59608 stars, 670 galaxies, and 562 YSOs  are labelled by c2d methods (see Table \ref{tab:num_of_source} for the number of data in different regions).

Finally, in order to prevent the bias or confusion due to the upper limit of fluxes for those non-detected bands, which results from the weak photon flux or extinction at shorter wavelengths, we replace these upper limits by 1/100 of the smallest reliable flux in that band. This is because even for non-detections, there are still thermal noises, which is in general much smaller than the reliable flux data. This replacement of upper limits by an artificial thermal noises allows us to utilize as much catalog data as possible without affecting the overall quality. We also have checked that different values of this artificial thermal noises (1/100 of the smallest reliable flux in that band) will not change the model training and test results if only small enough.

%=====================Table=======================%
\begin{table}
\centering
\begin{tabular}{c c c c}
\hline\hline
Region & Star & Galaxy & YSO\\
\hline
Chamaeleon II & 7516 & 34 & 26\\
ELAIS N1 & 398 & 49 & 0\\
Lupus I & 6783 & 43 & 14\\
Lupus III & 16567 & 32 & 74\\
Lupus IV & 5067 & 1 & 11\\
Ophiuchus & 17408 & 164 & 231\\
Perseus & 5869 & 347 & 206\\
\hline
Total & 59608 & 670 & 562 \\
\hline\hline
\end{tabular}
\caption{\label{tab:num_of_source}The regions and the number of sources labeled as stars, galaxies, and YSOs by \citet{Evans2007} after applying our data preparation, see Section \ref{sec:data_prep}}
\end{table}
%=================

%%%%%%%%%%%%
\subsection{Typical SEDs}

In Figure \ref{fig:SEDs}, we show some typical SEDs for our three types of objects: stars, galaxies, and YSOs, together with their errors for comparison. We can see that stars can be recognized by their black-body/gray-body radiation; the peak value is located in the short-wavelength region with an exponential tail extending to long-wavelength region. On the other hand, the SEDs for galaxies and YSOs are quite different from regular stars; their profiles usually reveal a double peak structure. The additional peak in the long-wavelength region is from the lower temperature dust surrounding the higher temperature stellar or protostellar objects. The relative amplitudes of these two peaks can vary in a wide range, making it difficult to distinguish them by SEDs alone.

In the lower row of Figure \ref{fig:SEDs}, we show the error distribution for different bands, after normalizing by the maximum value of their SEDs. One can see that in general the longest wavelength band, MIPS 1, has the largest errors compared to other bands. However, we will see later that this band may contribute most to the identification of YSOs.

%%%%%%%%%%%%%
\subsection{Flux-Error correlations}
\label{sec:Flux-Error}

We convert and compare the fluxes and errors of each band to a single regime to investigate whether these errors correlate to their corresponding fluxes in Figure \ref{fig:flux-error}.
We find that there are two trends in each panel of Figure \ref{fig:flux-error}. For (a)-(c), the sources located in the upper trend are observed by 2MASS, while those located in the lower trend are observed by UKIDSS. Since the UKIDSS observation has a longer exposure time compared to 2MASS, it gives a lower error for the same flux.
For (d)-(h), the observational data from the c2d legacy program and SWIRE legacy program are also shown together: 
the sources located in the upper trend are from the c2d legacy program, while the sources in the lower trend are from the SWIRE legacy program.
The observations from the SWIRE legacy program have a longer exposure time than the c2d legacy program, and therefore give lower error for the same flux.

\begin{figure*}[h]
\centering
\includegraphics[width=0.85\textwidth]{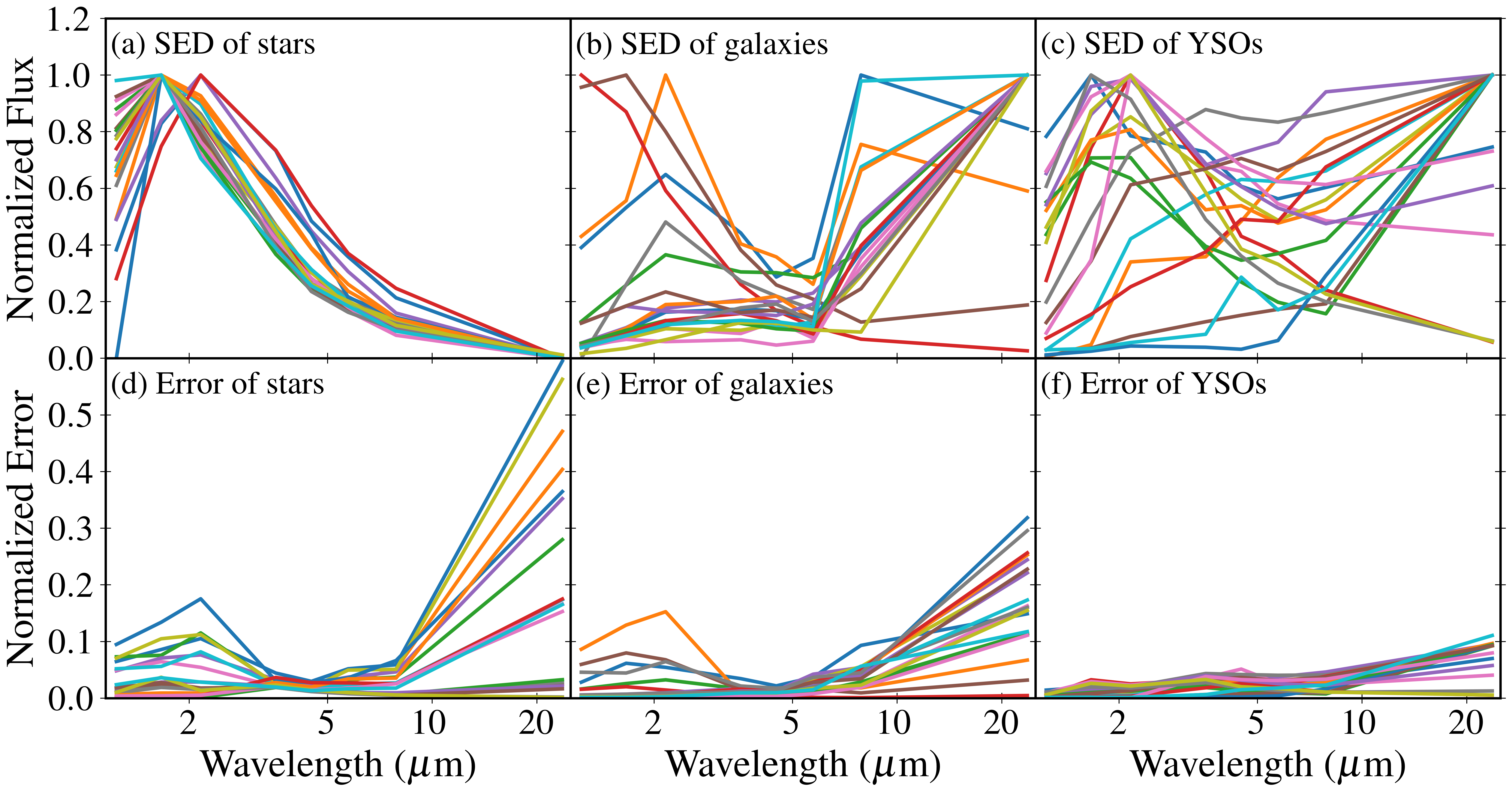}
\caption{Typical SEDs (upper row) and their errors (bottom row) for stars, galaxies and YSOs respectively (from left to right column). Note that, in order to show all results in the same scale, we have normalizied each SED and its error to the maximum value of SED. 
}
\label{fig:SEDs}
\end{figure*}

\begin{figure*}[h]
\centering
\includegraphics[width=0.85\textwidth]{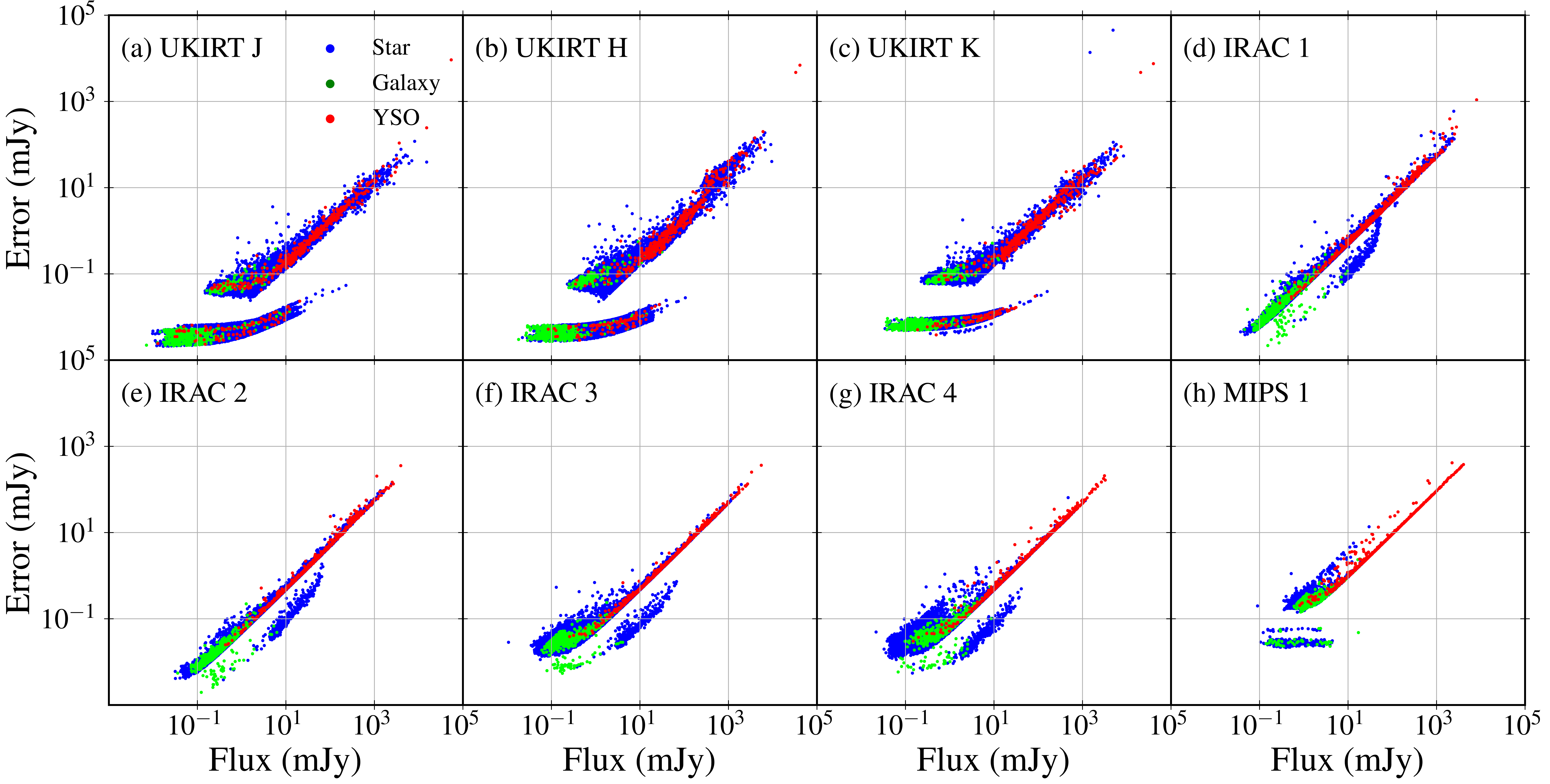}
\caption{Flux-Error plots for all of our data in the eight wavelengths. Red, green, and blue points represent stars, galaxies and YSOs, respectively. 
Each panel shows two groups of data:  
for (a)-(c), the data located in the upper/lower groups are obtained by 2MASS/UKIDSS;
for (d)-(h), the data located in the upper/lower groups are obtained by c2d legacy program/SWIRE legacy program. More details are described in Section \ref{sec:Flux-Error}.
}
\label{fig:flux-error}
\end{figure*}

The errors in the flux measurements mainly come from two origins: gain error of the instrument \citep{Evans2007} and thermal error. The former is almost a constant (depending on the observation time and mechanical limitation of the observatory), while the latter is from the statistical fluctuation of incoming photons and proportional to the flux when the flux becomes strong. As a result, the profile of the error distribution (see the lower row of Figure \ref{fig:SEDs}) also contains the flux information of these astronomical objects in each wavelength, although the relationship between these errors and the flux can differ between wavelengths and observatories. We there conclude that errors of each band may also provide additional information/features for the classification of astronomical objects as we will show below.
%==============
%=================
%===================
%%%%%%%%%%%%%%%%%%%%%%%%%%%%%%%%%%%%%%%%%%%%%%%
\section{Method} \label{sec:method}

In order to develop a proper machine learning model for the identification of stellar objects, we have used several typical classifiers for comparison, including Fully Connected Neural Network (FCN), eXtreme Gradient Boosting (XGB), Random Forest (RF), Support Vector Machine (SVM), and $k$ nearest neighbor (KNN) etc. 
These algorithms all work well and have similar performance. We decide to choose FCN as the major algorithm for SCAO, because it provides the highest recall and F1-score for YSO in all of the four models we design (see below). In other words, we expect that using FCN as the underlying algorithm could provide more accurate prediction of YSOs, and could be used for further studies by comparing results in different models. 

However, we also have to emphasize that the differences of performance between these models are quite close to each other, especially in Model I and Model II when all SED bands and errors are included. In Appendix \ref{Appendix: Comparison_ML} we will show the hyper-parameters and comparisons of these models for more details.

The artificial neural networks are computer algorithms originally designed to mimic the architecture of the biological brain in the 1940's \citep{ANN_early}, but was modified and improved to be so-far the most general and powerful scheme to realize artificial intelligence after the 2010's \citep{ANN}. Its common structure is composed of one input layer and one output layer, where in between are multiple hidden layers. In each layer, there are a number of interconnected nodes, referred to as ``neurons", which receive inputs from other ``neurons" in a previous layer and supply outputs to the ``neurons" in the next layer. Each node performs a weighted sum computation on the values it receives from the input and then generates an output using a simple nonlinear transformation function on the summation. 

To carry out the classification of these stellar objects, we express our data as 1D arrays with 16 features, meaning 8 bands with flux measurement and error (i.e. the Model II in SCAO). Our FCN model consist of two hidden layers with 50 neurons, which is fixed for different types of input data in Section \ref{sec:results}, and one output layer. An activation function for interpreting and transmitting information between layers is chosen to be an Scaled Exponential Linear Unit (SELU) function. The output is normalized to the probability by using the softmax function. The loss function and the optimizer are set to cross-entropy and adam \citep{Adam}, which are the common choices in machine learning application. 

As for the training process, we divide our data into three subsets, 70\% for training, 10\% for validation and 20\% for test. The validation set is used to evaluate which parameters could give the best performance during the training process, and the test set is used to show the performance for the unknown data. The neuron networks are trained for more than 280 epochs in each run in order to ensure the full saturation of training accuracy ($>$ 99.99\% when applied to the validation set). In the whole training process, the model with highest F1-score of YSO candidate on the training set and validation set will be selected as our classifier. We also have checked carefully the potential over-fitting problem during the training process, and have taken the proper hyper-parameters to remove it in all the models and calculations shown below.

Finally, we train each model with ten complete runs independently, and then use the softmax function to average the ten probabilities as the final output result. This can help stabilize our results and improve the reliability for future applications. 

%We note that, since the number of data points in our classifier, SCAO, are limited by the energy bands available for each source (see \autoref{tab:bands}), SCAO is mainly designed to classify the mutual correlation of these SEDs and their error patterns, wherein physical origins might be hidden. This is different from the ordinary pattern recognition problem and therefore certain rule-based or statistics-based machine learning algorithms may still work as well as those neural networks.

%====================
\begin{table}
\centering
\begin{tabular}{cc|c|c|c|}
& \multicolumn{1}{c}{} & \multicolumn{3}{ c }{predicted labels} \\ 
\cline{2-5}
& \multicolumn{1}{|c|}{} & {\bf Star} & {\bf Galaxy} & {\bf YSO}  \\ 
\cline{2-5}
\multicolumn{1}{ c  }{actual} &
\multicolumn{1}{ |c| }{{\bf Star}} & {\it SS} & {\it SG} & {\it SY}  \\ 
\cline{2-5}
\multicolumn{1}{ c  }{labels}                        &
\multicolumn{1}{ |c| }{{\bf Galaxy}} & {\it GS} & {\it GG} & {\it GY}  \\ 
\cline{2-5}
\multicolumn{1}{ c  }{}                        &
\multicolumn{1}{ |c| }{{\bf YSO}} & {\it YS} & {\it YG} & {\it YY} \\
\cline{2-5}
\end{tabular}
\caption{The Confusion matrix for the three types of astronomical objects.}
\label{tab:confusion matrix}
\end{table}
%================

%%%%%%%%%%%%%%%%%%%%%%%%%%%%%%%%%%%%%%%%%%%%%%%%%%%%%%%%%%%%%%%%%%%%%%
\section{Results} 
\label{sec:results}

In this section, we show results obtained by different models SCAO based on four different types of input features: Model I, using SEDs only; Model II, using SEDs with errors; Model III, using normalized SEDs; Model IV, using SEDs only containing data from the three bands in the longest wavelength regimes. These models have different purpose for research: Model I and Model II are aimed to show that SCAO can be used to classify objects as proofs of concept. Model III is used to check whether sources can be classified without any distance information. Model IV is aimed to check whether YSOs can be successfully identified with fewer bands.

In order to understand the performance of our four SCAO models, we apply them to the test data to obtain confusion matrices, where the predicted results are compared with their actual labels (see Table \ref{tab:confusion matrix}). We have to emphasize here that the ``actual label" is the label obtained by the c2d method \citep{Evans2007}, which is based on some existing theories or assumptions as described in Section \ref{sec:actual_labels}. They may not be strictly correct if further observations are obtained by better telescopes. However, since this paper is to develop a general model based on data alone, we will still assume that these "actual labels" are correct and will not discuss their validity here.

The confusion matrices in our paper are defined in Table \ref{tab:confusion matrix}, where each label in the middle (SS, SG, GY etc.) is the number of objects. For example, SY means the number of objects with an ``actual label" to be a star, but predicted to be a YSO. In other words, the off-diagonal terms of this confusion matrix denote the difference between ``actual labels" and ``predicted labels". From this confusion matrix, the precision and recall can be defined for each type of objects. For example, the precision ($P$) and recall ($R$) of YSOs are respectively defined as
%======================
\begin{equation}
    \begin{aligned}
    P_{YSO} &=& \frac{YY}{SY+GY+YY} \\
    \end{aligned}
\end{equation}
\begin{equation}
    \begin{aligned}
    R_{YSO} &=& \frac{YY}{YS+YG+YY} \\
    \end{aligned}
\end{equation}
\begin{equation}
    \begin{aligned}
    F1_{YSO} &=& \frac{2P_{YSO}R_{YSO}}{P_{YSO}+R_{YSO}}.
    \end{aligned}
\end{equation}
%=========
Similar definitions can be also obtained for star ($P_{Star}$, $R_{Star}$, and $F1_{Star}$) and galaxy ($P_{Galaxy}$, $R_{Galaxy}$, and $F1_{Galaxy}$) respectively. Precision means the percentage to get accurate result out of all the predicted events, while recall is the percentage to be predicted from the known objects. F1 score is the harmonic mean of the precision and recall. These quantities are correlated to each other. The overall accuracy for a given model is defined by the ratio of diagonal terms with respect to all data:
%===============
\begin{eqnarray}
Acc &=& \frac{SS+GG+YY}{SS+SG+\cdots+YG+YY}.
\end{eqnarray}
%============
\begin{widetext}
\centering*
\begin{table}
\begin{tabular}{|c||c|c|c|c|c|c||c|c|c|c|c|c||c|c|c|c|c|c||c|c|c|c|c|c|}
\hline
Model & \multicolumn{6}{|c||}{Model I (99.8\%)} & \multicolumn{6}{|c||}{Model II (99.9\%)} & \multicolumn{6}{|c||}{Model III (99.5\%)} & \multicolumn{6}{|c|}{Model IV (99.7\%) } \\
\hline
Data & \multicolumn{6}{|c||}{SED} & \multicolumn{6}{|c||}{SED with error} & \multicolumn{6}{|c||}{Normalized SED} & \multicolumn{6}{|c|}{Partial SED} \\
\hline
 & \multicolumn{2}{|c|}{S} & \multicolumn{2}{|c|}{G} & \multicolumn{2}{|c||}{Y} & \multicolumn{2}{|c|}{S} & \multicolumn{2}{|c|}{G} & \multicolumn{2}{|c||}{Y} & \multicolumn{2}{|c|}{S} & \multicolumn{2}{|c|}{G} & \multicolumn{2}{|c||}{Y} & \multicolumn{2}{|c|}{S} & \multicolumn{2}{|c|}{G} & \multicolumn{2}{|c|}{Y} \\ 
\hline
S & \multicolumn{2}{|c|}{11911} & \multicolumn{2}{|c|}{5} & \multicolumn{2}{|c||}{5} & \multicolumn{2}{|c|}{15918} & \multicolumn{2}{|c|}{2} & \multicolumn{2}{|c||}{1} & \multicolumn{2}{|c|}{15894} & \multicolumn{2}{|c|}{15} & \multicolumn{2}{|c||}{12} & \multicolumn{2}{|c|}{11907} & \multicolumn{2}{|c|}{7} & \multicolumn{2}{|c|}{7} \\ 
\hline
G & \multicolumn{2}{|c|}{1} & \multicolumn{2}{|c|}{132} & \multicolumn{2}{|c||}{1} & \multicolumn{2}{|c|}{1} & \multicolumn{2}{|c|}{130} & \multicolumn{2}{|c||}{3} & \multicolumn{2}{|c|}{1} & \multicolumn{2}{|c|}{125} & \multicolumn{2}{|c||}{8} & \multicolumn{2}{|c|}{1} & \multicolumn{2}{|c|}{130} & \multicolumn{2}{|c|}{3} \\ 
\hline
Y & \multicolumn{2}{|c|}{3} & \multicolumn{2}{|c|}{1} & \multicolumn{2}{|c||}{108} & \multicolumn{2}{|c|}{1} & \multicolumn{2}{|c|}{1} & \multicolumn{2}{|c||}{110} & \multicolumn{2}{|c|}{2} & \multicolumn{2}{|c|}{13} & \multicolumn{2}{|c||}{97} & \multicolumn{2}{|c|}{1} & \multicolumn{2}{|c|}{3} & \multicolumn{2}{|c|}{108} \\ 
\hline
\hline
& \multicolumn{3}{|c|}{Precision} & \multicolumn{3}{|c||}{Recall}
& \multicolumn{3}{|c|}{Precision} & \multicolumn{3}{|c||}{Recall}
& \multicolumn{3}{|c|}{Precision} & \multicolumn{3}{|c||}{Recall}
& \multicolumn{3}{|c|}{Precision} & \multicolumn{3}{|c|}{Recall} \\
\hline
S & \multicolumn{3}{|c|}{99.97\%} & \multicolumn{3}{|c||}{99.92\%}
& \multicolumn{3}{|c|}{99.98\%} & \multicolumn{3}{|c||}{99.97\%}
& \multicolumn{3}{|c|}{99.97\%} & \multicolumn{3}{|c||}{99.77\%}
& \multicolumn{3}{|c|}{99.98\%} & \multicolumn{3}{|c|}{99.88\%} \\
\hline
G & \multicolumn{3}{|c|}{95.7\%} & \multicolumn{3}{|c||}{98.5\%}
& \multicolumn{3}{|c|}{97.7\%} & \multicolumn{3}{|c||}{97.0\%}
& \multicolumn{3}{|c|}{81.5\%} & \multicolumn{3}{|c||}{93.3\%}
& \multicolumn{3}{|c|}{92.7\%} & \multicolumn{3}{|c|}{97.0\%} \\
\hline
Y & \multicolumn{3}{|c|}{94.8\%} & \multicolumn{3}{|c||}{96.4\%}
& \multicolumn{3}{|c|}{96.5\%} & \multicolumn{3}{|c||}{98.2\%}
& \multicolumn{3}{|c|}{82.9\%} & \multicolumn{3}{|c||}{86.6\%}
& \multicolumn{3}{|c|}{91.5\%} & \multicolumn{3}{|c|}{96.4\%} \\
\hline
\end{tabular}
\centering*
\caption{Summary of the confusion matrix and precision/recall table for the four different models in our SCAO, trained by four different types of datasets (see the text). For each model, we show the overall accuracy in the top row, and calculate all results with averaged FCN parameters, which are obtained from 10 independent training processes. S, G, and Y indicate star, galaxy, and YSO, respectively. The test data contain 11921 stars, 134 galaxies, and 112 YSOs, about 20\% of the total objects, see Table \ref{tab:num_of_source}.
}
\label{tab:confusion matrix for all}
\end{table}
\end{widetext}
%===================
%%%%%%%%%%%%%%%%%
\subsection{Model I: Classification using SED}
\label{sec:M1_proof}

We first show the calculated results of Model I, which is trained and tested by using only the SEDs, that is, there are eight data points (for the 8 different bands) in total for each astronomical object. The corresponding telescopes and wavelengths are shown in Table \ref{tab:bands}. We note that, although their absolute flux amplitudes reflect their distance to the Earth, their relative amplitude, that is, the profile/shape of the SED structure, contains the true physical information. In the first column of Table \ref{tab:confusion matrix for all} we show the confusion matrix for the predicted results, compared to their actual labels according to previous theory \citep{Evans2007}. 

In the lower part of this column, we show the calculated precision and recall for each type of these objects. One could see that results for stars are extremely good: both precision and recall are higher than 99\%, showing that the standard SED pattern for a regular star, has been learned and correctly captured, even though we did not implement any theoretical equations or assumptions in our model. As for galaxy and YSO labels, although both of them can be recalled very well ($>96$\%), their precision are smaller by several percentages ($\sim 95$\%). The overall accuracy of Model I is 99.8\%, due to the high precision prediction of stars 

%The differences in precision and recall can be understood from the first row of their confusion matrix, where 5 stars are predicted to be galaxies and 5 stars predicted to be YSO. These numbers are several times more than the reverse cases. Although, these are a very small amount compared to the total number of stars, but can seriously affect the prediction of galaxy and YSO, for their total numbers are not as much as stars. We will see later that such discrepancy can be recovered when the observation errors are included.
%%%%%%%%%%%%%%%%%
\subsection{Model II: Classification using SED with error}

Since the classification process of the c2d method includes errors, it is also necessary to see if SCAO could provide better results than Model I, when the observation errors are included. As a result, we train Model II
with the SEDs and their observational errors (i.e. there are sixteen data points in total for each astronomical object) for comparison. The obtained confusion matrix as well as the precision/recall are shown in the second column of Table \ref{tab:confusion matrix for all}.

Naively speaking, measurement uncertainties (errors) seem to have nothing to do with the nature of the light sources. As shown in Figure \ref{fig:flux-error} and explained in Section \ref{sec:Flux-Error}, the observation uncertainty can be actually correlated to the amplitude of flux through simple photon statistics and therefore could be also considered as additional data input. If the error bar of a measurement is comparable to or even larger than its mean value, it indicates that the measured mean value (i.e. the flux) is not reliable and one may interpret the whole SED differently for the classification (for example, a galaxy may be identified as a star or vice versa). 
This is indeed also what happens in the classification of the c2d method \citep{Evans2007}, and therefore has been learned and recovered in our Model II through the input data.

We can see that when we include errors in the training process, the confusion matrix becomes almost diagonal, meaning the misprediction of stars, galaxies, and YSOs become much less, and therefore the precision and recall become better than in Model I (Table \ref{tab:confusion matrix for all}). This result shows that our SCAO could successfully recover almost all the results obtained by the c2d legacy project, which was guided by some theoretical assumptions \citep{Evans2009}. 
We apply Model II on sources extracted from SEIP catalog, and obtain 45724 YSOs from 0.66M SEIP sources using the same criteria described in Section \ref{sec:data_prep}. Our YSO candidate list is shown in Appendix \ref{Appendix: MRT for SEIP}.

%%%%%%%%%%%%%%%%%
\subsection{Model III: Classification using normalized SED} 

In the third column of Table \ref{tab:confusion matrix for all}, we show the confusion matrix as well as precision/recall obtained by Model III, which is trained and tested by using the SEDs normalized by their maximum value without the error, that is, there are eight data points (for the 8 different bands with flux measurement) in total and one constrain for each astronomical object.
Different from Model I, which uses the original SED flux without normalization, the motivation of Model III is to remove the distance effect from the overall flux amplitude and investigate whether our SCAO model could differentiate the SED pattern of a galaxy from that of a YSO. Due to the limitation of observation sensitivity, all the observable stars or YSOs should be within our Galaxy, while all galaxies are certainly much further away. In other words, their SEDs may be distinguished simply based on their absolute brightness, even though their profiles are similar to each other (see Figure \ref{fig:SEDs}). After all, there is still some possibilities that fainter stars/YSOs or brighter galaxies can be confused.

In the resulting confusion matrix shown in the third column of Table \ref{tab:confusion matrix for all}, we can see that the off-diagonal elements are more than their values obtained by SEDs only in Model I. The precision and recall of galaxy and YSO are therefore reduced as expected due to their similar profiles. However, we surprisingly find that SCAO could still get a good recall ($>86$\%) to distinguish YSOs from galaxies and stars. In other words, even though the distance effect has been completely removed in the normalized SEDs, there may still exists some {\it hidden} structures distinguishable by our SCAO model that could be critically important for classifications if the SEDs are too similar to each other.
Figuring out how to extract these hidden structures in normalized SEDs can provide further information for understanding the star formation process. 
%=========================
\begin{figure*}[htb!]
\centering
\includegraphics[width=0.96\textwidth]{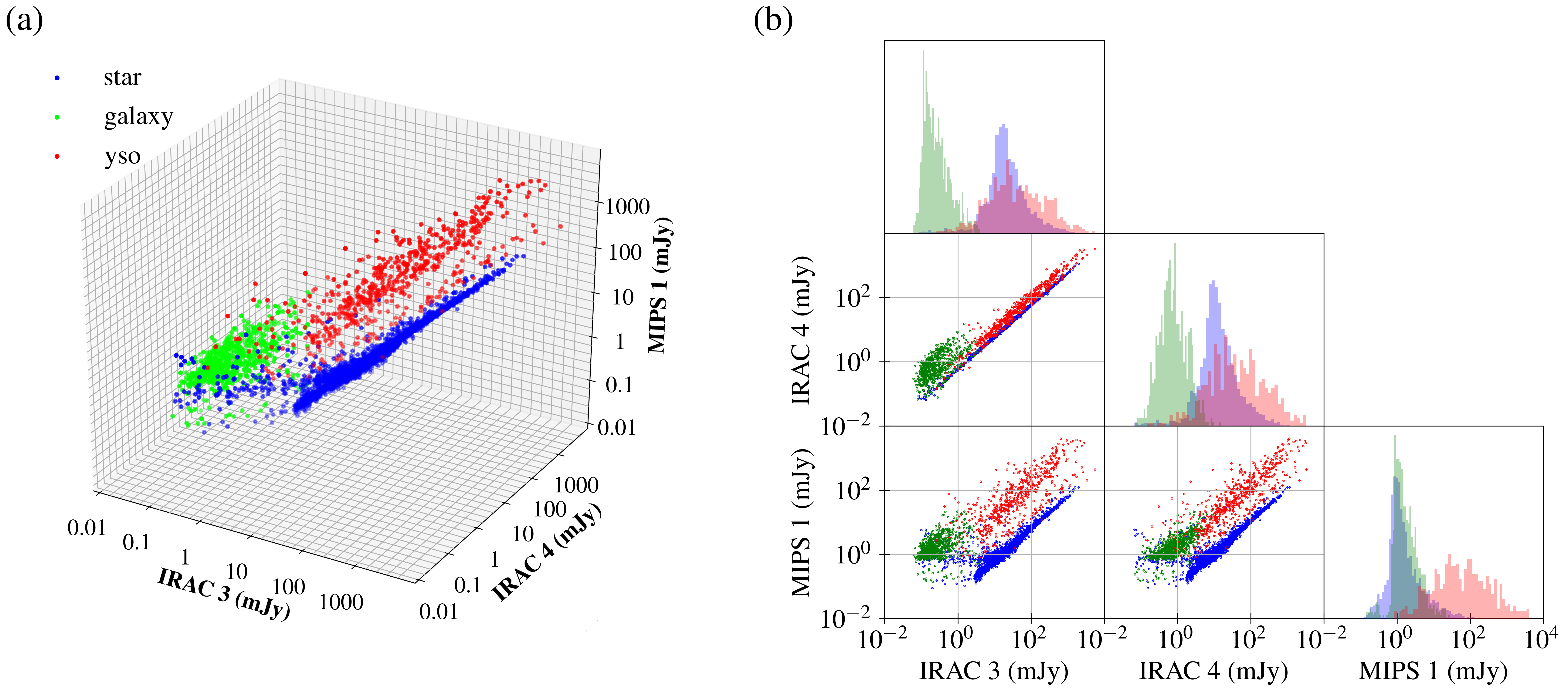}
\caption{
(a) The 3D log-log-log plot of all data in the last three SED bands (IRAC 3, IRAC 4, and MIPS 1). Red, Green and Blue dots represent Stars, Galaxies, and YSOs respectively. One could see that these three types of objects are quite well separated in such 3D space, showing why data from these three bands play the most important roles to distinguish them, even though SEDs of shorter wavelengths are not included at all.
(b) The corner plots obtained by the same three SED bands, showing a strong overlap of these objects in any projection of a two dimensional plane. 
}
\label{fig:SED 3D}
\end{figure*}
%=========================
%%%%%%%%%%%%%%%%%
\subsection{Model IV: Classification using partial SED}
\label{sec:M4_partial_SED}

Although the performance of our Model I and Model II are very good, reproducing almost all the results obtained by c2d method, their application are still limited by the fact that most of the sources are not full-band detected. In order to extend the application regime, it is reasonable to check if we could get a reasonably good performance from a model with fewer number of bands.  
We have tried all possible three band combinations, and found that the best combination is obtained by SEDs of IRAC 3, IRAC 4 and MIPS 1, the three bands in the longest wavelength regime in our database. 
Thus, we conclude that IRAC 3, IRAC 4 and MIPS 1 provide the most important information for identifying YSOs, and we construct Model IV trained with the data from these three bands.

In the last column of Table \ref{tab:confusion matrix for all}, we show the confusion matrix as well as precision/recall obtained by Model IV, which is trained and tested by using the SEDs only containing data points from the three bands in the long wavelength regime (IRAC 3, IRAC 4, and MIPS 1), that is, there are three data points (for the 3 different bands with 1 flux measurement) in total for each astronomical object. The obtained precision for predicting YSOs can be above 96\%, which is equal to the results using the full (eight) SED bands in Model I. On the other hand, the precision and recall for the other two types (star and galaxy) are reduced only slightly.
This result may also reflect the fact that the extinction effect is stronger at shorter wavelengths, making the FCN model confused when including the whole eight band SED. 

In order to visualize above results more precisely, in Figure \ref{fig:SED 3D}(a) we show a 3D plot for the fluxes of these three bands. We can see that these three types of objects are well separated in 3D space, and can therefore be easily classified by SCAO. However, their distribution cannot be easily separated if we project onto the axis of a single wavelength (see Figure \ref{fig:flux-error}) or onto the plane of two wavelengths (see Figure \ref{fig:SED 3D}(b)).

In fact, we have also checked the results using the three bands in the short wavelength regime, but could not achieve a precision and recall above 80\%. Our results strongly suggest that for the observation of YSOs, the SED features from at longer wavelengths ($5.7\mu{\rm m}<\lambda<24.0\mu{\rm m}$), where the major contribution is dominated by the dust around the stellar cores, are much more important than those at shorter wavelengths, even though the observational uncertainty of the latter is also much smaller. An obvious and reasonable explanation is because of the well-known silicate absorption within this wavelength regime \citep{Draine2003}, which is a strong evidence for a YSO source. 
Although this feature (silicate absorption) is not directly observed in the present broad-band observations, our SCAO in Model IV could still capture the hidden information of such signature of YSOs based on the broad band spectrum at IRAC 3, IRAC 4, and MIPS 1. We also have applied Model IV on sources extracted from the SEIP catalog, and obtain more than 0.12M YSOs from 1.4M sources. Our results imply that using three band only (Model IV) could still find much more YSO candidates with reasonable accuracy, making SCAO more useful than other models.

%We require sources that were observed in IRAC 3, IRAC 4, and MIPS 1 bands, and detected in IRAC 3 and IRAC 4 bands,
%i.e. non-detections in MIPS 1 is allowed. We then assign a flux value of 1/100 of the smallest reliable flux to non-detections in that band. The list of prediction result is shown in \autoref{Appendix: MRT for SEIP}.

%%%%%%%%%%%%%%%%%%%%%%%%%%%%%%%%%%%%%%%%%%%%%
\section{Discussion}
\label{Sec:discussion}

%%%%%%%%%%
\subsection{Extinction effects}

In the model training and test results shown above, we just use observed flux and/or errors as the input features. However, since many sources in the c2d catalog are embedded in or behind molecular clouds, it is necessary and instructive to discuss the extinction effects in our results. After all, it is known that extinction effects would change the shapes of the SEDs, since dust grains in molecular clouds absorb more emission at shorter wavelengths.

In order to know whether the extinction effect biases our predictions, we use the extinction values given by Near-Infrared Color Excess Revisited technique \citep[NICER;][]{Lombardi2001, Meingast2017} and the observed fluxes of our sources to calculate their intrinsic fluxes, to create a new dataset. We then train and test our SCAO model on this dataset, and compare to the results of Model II in Table \ref{tab:confusion matrix for all}.

We find that the predicted results of this model are almost the same as those given by Model II (to save space and to avoid confusion, we do not show them here), except that there are 17 sources predicted differently than in the c2d method. 
The overall accuracy is 99.90\% with the YSO precision and recall to be  94.2\% and 94.8\% respectively, slightly lower than the results of Model II in Table \ref{tab:confusion matrix for all}. This indicates that the most important contribution of SEDs for identifying YSO are those bands in the long wavelength regime where the extinction effect is negligible (for example, the success of Model IV uses only three bands of the longest wavelengths). 
As a result, we could conclude that extinction effects should not be significant in the application of SCAO. 
%%%%%%%%%%%

%======
\begin{table}
\centering
\begin{tabular}{c|c| c c c c}
    \hline\hline
    Regions  & Model & Star & Galaxy & YSO & Error Rate\\
    \hline
    
    \multirow{3}{*}{CDFS} &  II & 442 & 843 & 6 & 0.46\%\\
    & IV & 12591 & 19037 & 13 & 0.04\% \\
    \hline
    
    \multirow{3}{*}{ELAIS N2} & II & 257 & 412 & 9 & 1.33\% \\
    & IV & 4202 & 8348 & 13 & 0.10\% \\
    \hline
    
    \multirow{3}{*}{ELAIS S1} & II & 117 & 505 & 6 & 0.96\% \\
    & IV & 2965 & 7326 & 12 & 0.12\% \\
    \hline
    
    \multirow{3}{*}{Lockman} & II & 715 & 1123 & 8 & 0.43\% \\
    & IV & 12717 & 22379 & 21 & 0.06\% \\
    \hline
    
    \multirow{3}{*}{XMM-LSS} & II & 479 & 1180 & 11 & 0.66\%\\
    &  IV & 10865 & 17105 & 24 & 0.09\% \\
    \hline
    
    \multirow{3}{*}{Total} & II & 2010 & 4063 & 40 & 0.65\% \\
    & IV & 43340 & 74195 & 83 & 0.07\% \\
    \hline\hline
    
\end{tabular}
\caption{The SCAO predicted numbers of stars, galaxies and YSOs in SWIRE-defined extra-galactic regions, where there should be no YSO at all. The last percentage shows the errors of SCAO. Note that the total number of objects classified by Model IV are more than those by Model II, because the latter requires SED observation of full 8 bands.  
\label{tab:pred_swire}}
\end{table}
%============
%============
\begin{table}
\centering
\begin{tabular}{c|c| c c c c}
    \hline\hline
    Regions & Model & Star & Galaxy & YSO & Recall of YSO\\
    \hline
    An et al. &  II & 0 & 0 & 23 & 100.0\%\\
    (2011) & IV & 0 & 0 & 28 & 100.3\% \\
    \hline
    Furlan et &  II & 0 & 1 & 100 & 99.0\%\\
    al. (2016)&  IV & 3 & 11 & 272 &  95.1\%\\
    \hline\hline
\end{tabular}
\caption{The SCAO predicted number of stars, galaxies and YSOs on spectroscopically-confirmed YSOs. Note that the total number of objects classified by Model IV are more than those by Model II, because the latter requires SED observation of full 8 bands.  
\label{table: pred_spec_YSO}}
\end{table}
%=============
%%%%%%%%%%%%%%%%%%%%%%%%%%
\subsection{Validation of SCAO}
\label{sec:val_SCAO}

Although we have shown that SCAO can reproduce almost all of the classifications given by the c2d method and can be extended to more sources by using three bands only in the mid-infrared region, it is helpful to compare the predicted results from those known sources with the ground truth.
Here we consider two independent approaches for the comparison of YSO prediction. 

We first apply SCAO to classify astronomical objects using SEIP sources located in five SWIRE-defined extragalactic regions, which are cloudless and should have no YSOs (Table \ref{tab:pred_swire}). 
ELAIS N1 region is not included since we train our models with sources from there.
In the last column of Table \ref{tab:pred_swire} we show the error rate of SCAO model predictions for YSOs. 
We find the error rate of Model II is in general less than 1\%, while the error rate of Model IV is even lower than 0.1\%.
One of the possible reason is that Model IV requires only three bands for the classification and therefore could be applied to many more sources. This result shows that our SCAO, especially Model IV, can have extremely good precision in the search for YSOs. 

Additionally, we apply SCAO to classify spectroscopically confirmed YSOs obtained from \citet{An2011} and \citet{Furlan2016} (\autoref{table: pred_spec_YSO}). Our results show that all YSOs in \citet{An2011} and 96.15\% of them in \citet{Furlan2016} are correctly identified. This implies our SCAO also has a very good recall rate on YSOs. Both two results support our previous claim that SCAO does learn some essential aspects for the classification of YSOs and the high precision/recall are independent of source region or training dataset. Moreover, our Model IV with only three bands (IRAC 3, IRAC 4 and MIPS 1) can certainly be applied to more sources than Model II with good accuracy even when compared to the known ground truth.

%%%%%%%%%%%%%%
\subsection{Comparison of the classification results with \citet{Marton2019}}

So far, we have shown that our SCAO is well-trained and tested not only withing our dataset, but also compared with the ground truth for both cloudless regimes (no YSOs) and for known YSOs. Nevertheless, it is important to check how our results are consistent with results using similar machine learning methods. To achieve this purpose, we compare the predictions of our SCAO to the all-sky-source predictions provided in \citet{Marton2019}, where the authors constructed two supervised random forests, model L and model S, to classify infrared sources using fluxes from certain bands. Model L uses fluxes of bands ranging from visible light to middle infrared, include G band in Gaia DR2, JHK bands in 2MASS, and 4 WISE bands in ALLWISE. Model S uses fluxes of the same bands of model L except WISE 3 and WISE 4, in order to get rid of possible spurious detections \citep{Marton2016, Marton2019, Koenig2014}. We note that the definition of our source type ``star" actually includes both main-sequence stars (MS) and evolved stars (ES), and therefore we will add the numbers of MS and ES provided by \citet{Marton2019}.

We first compare the predictions of our Model II with Model L on sources of SEIP catalogs, since they have similar wavelength coverage, and the results are in Table \ref{tab:IIvsMarton}. As described above, we add numbers of MS and ES together for comparison. One can see that there are obvious discrepancies in all three types of objects. We speculate that the disagreement between Model L and Model II mainly comes from the fact that \citet{Marton2019} takes almost all available YSO candidate lists for the training process from all YSO-related \textit{Spitzer} publications and the SIMBAD database.
It is well-known that different methods have different standards/criteria on classifications and could reduce the internal consistency.
Also, the c2d method we adopted for our training labels is known to be one of the most restrictive methods, since it excludes sources with any possibility to be stars or galaxies. 

In order to understand the internal consistency of these two approaches, we show the confusion matrix between our Model II and Model IV in Table \ref{tab:IIvsIV}. The overall consistency can be calculated from the ratio of the diagonal terms (correctly predicted by both Models) to the summation of total number of objects. We find it is as high as 96.8\%. This value is reduced to 93.8\% when comparing Model L and Model S provided in \citet{Marton2019} (see also, Table \ref{tab:LvsS}). 
Furthermore, the YSO recall and precision of Model L are 93.5\% and 90.68\%, respectively. Both are lower than Model II by 5\%. The recall and precision of Model S and Model IV are comparable, with 93.14\% recall and 91.78\% precision for Model S.

As a result, after the above comparison between different models in SCAO and in \citet{Marton2019}, we conclude that Model II and Model IV of our SCAO are more consistent with each other than the Model L and Model S. This reflects the fact that c2d labels should be more consistent and restricted within itself and hence better for the training of machine learning model to identify YSO.  

%======
\begin{table}
\centering
\begin{tabular}{|c|c|c c| c c c|}
    \hline\hline
      & & \multicolumn{5}{c|}{Model L} \\
     \hline
      & & MS & ES  & MS+ES & G  & Y \\
     \hline
     \multirow{3}{*}{\rotatebox[origin=c]{90}{Model II}} 
     & S & 3639 & 225918 & 229557 & 881 & 24612\\
     & G & 40 & 662 & 702 & 276 & 1802\\
     & Y & 61 & 5935 & 5996 & 80 & 13710\\
    \hline\hline
\end{tabular}
\caption{The confusion matrix comparing the classifications of Model II to Model L in \citet{Marton2019}. Here MS indicates ``main-sequence stars", ES indicates ``evolved stars", and MS+ES indicates the total number of main-sequence stars and evolved stars.
\label{tab:IIvsMarton}}
\end{table}

%=========
\begin{table}
\centering
\begin{tabular}{|c|c|c c c|}
    \hline\hline
      & & \multicolumn{3}{c|}{Model IV} \\
     \hline
      & & S & G & Y \\
     \hline
     \multirow{3}{*}{\rotatebox[origin=c]{90}{Model II}} 
     & S & 255002 & 23 & 25 \\
     & G & 747 & 1463 & 570 \\
     & Y & 3916 & 74 & 15796 \\
    \hline\hline
\end{tabular}
\caption{ The confusion matrix comparing the classifications of Model II to Model IV.
\label{tab:IIvsIV}}
\end{table}
%=========
%=========
\begin{table}
\centering
\begin{tabular}{|c|c|c c|c c c|}
    \hline\hline
      & & \multicolumn{5}{c|}{Model S} \\
     \hline
      & & MS & ES & MS+ES & G & Y \\
     \hline
     \multirow{4}{*}{\rotatebox[origin=c]{90}{Model L}} 
     & MS & 3189 &   118 & 3307 & 2 & 431\\
     & ES & 141 & 224092 & 224233 & 285 & 7997\\
     \cline{2-7}
     & MS+ES & 3330 &  224210 & 227540 & 287 & 8428\\
     & G & 1 &  61 & 62 & 1052 & 123\\
     & Y & 42 &  8038 & 8080 & 111 & 31933\\
    \hline\hline
\end{tabular}
\caption{ The confusion matrix comparing the classifications of Model L to Model S. 
\label{tab:LvsS}}
\end{table}

%%%%%%%%%%%%%%%%
%\subsection{Predicted distribution regime in the flux space.}

%In Sec.\,\ref{sec:M4_partial_SED}, we have shown that our Model IV could provide accurate predictions for stars, galaxies and YSOs by using only three band SEDs (IRAC 3, IRAC 4 and MIPS 1). This can be clearly understood by showing the distribution of all sources in our dataset in a 3D flux space (Fig. \ref{fig:SED 3D}). Based on Model IV, we can determine whether a source is a YSO simply by their IRAC 3, IRAC 4, and MIPS 1 fluxes. Fig. \ref{fig:prob_dist_Model_IV} shows the minimum MIPS 1 flux required to be a YSO with given IRAC 3 and IRAC 4 fluxes. Although this plot could not fully exclude the tiny possibility to find a YSO with MIPS1 flux lower than the "minimum" value since the accuracy is not 100\%, it is still a convenient tool for new observations and an important benchmark for  future theoretical works. 

%%=============
%\begin{figure}[htb!]
%\centering
%\includegraphics[width=0.5\textwidth]{mcm_contour_probability_for_all_sources.png}
%\caption{
%The minimum MIPS 1 fluxes of sources being YSO using Model IV. Red dots indicate the YSOs we use in training process (Sec. \ref{Sec:data_prep}).
%}
%\label{fig:prob_dist_Model_IV}
%\end{figure}
%%%%%%%%%%%%%%%%%%%%%%%%%%%%%%%%
\section{Conclusion}

In this paper, we construct a new machine classifier, SCAO, to classify three major astronomical objects: stars, galaxies, and YSOs, based on their SEDs through Fully Connected Neural Network. Different from previous classifiers, our model is trained by labeled data without any theoretical knowledge or assumptions, and provides excellent results with very high precision and recall. The classification is trained from four different kinds of input data in order to satisfy different datasets for different research purposes (see Table \ref{tab:confusion matrix for all}).
By investigating the classification results of normalized SEDs, we demonstrate that YSOs can still be well differentiated from galaxies even when the distance effects are completely removed. 
We show that the major features comes from the flux in the longer infrared wavelengths, reflecting the importance of dust components. 
To test the performance of SCAO, we apply our models on sources located in extragalactic regions, and the number of incorrectly predicted YSOs is negligibly small. On the other hand, our models can also recover more than 95\% of the YSOs from two confirmed YSO lists.
We also compare our model II to the model L of \citet{Marton2019}, and find that the consistency between the two models is $\sim$86\%;
we speculate that our model trained with a single method for source identifications would have higher internal consistency.
Our results show that the pattern recognition of SED structure can be applied to differentiate YSOs from galaxies and stars, and provide an independent approach to investigate the physical mechanism behind the existing star formation theory. 
We provide a website to allow people to use SCAO models and classifications easily at http://scao.astr.nthu.edu.tw.

%%%%%%%%%%%%%%%%%%%%%%%%%%%%%%%%%%%%
%%%%%%%%%%%%%%%%%%%%%%%%%%%%%%%%%%
{\bf Acknowledgement}

YLC and SPL acknowledge support from the Ministry of Science and
Technology of Taiwan with Grant MOST 106-2119-M-007-021-MY3.
CTH and DWW acknowledge support from National Center for Theoretical Sciences and from the Ministry of Science and Technology of Taiwan with Grant MOST 107-2112-M-007-019-MY3.
This work is also supported by the Higher Education Sprout Project funded by the Ministry of Science and Technology and Ministry of Education in Taiwan.
This work is based on observations made with the \textit{Spitzer} Space Telescope, which is operated by the Jet Propulsion Laboratory, California Institute of Technology under a contract with NASA.
% for 2MASS
This publication makes use of data products from the Two Micron All Sky Survey, which is a joint project of the University of Massachusetts and the Infrared Processing and Analysis Center/California Institute of Technology, funded by the National Aeronautics and Space Administration and the National Science Foundation.
% for UKIRT
UKIRT is owned by the University of Hawaii (UH) and operated by the UH Institute for Astronomy; operations are enabled through the cooperation of the East Asian Observatory.
When (some of) the data reported here were acquired, UKIRT was supported by NASA and operated under an agreement among the University of Hawaii, the University of Arizona, and Lockheed Martin Advanced Technology Center; operations were enabled through the cooperation of the East Asian Observatory.
% for Planck Legacy Archive
Based on observations obtained with Planck (http://www.esa.int/Planck), an ESA science mission with instruments and contributions directly funded by ESA Member States, NASA, and Canada.
% for SIMBAD
This research has made use of the SIMBAD database, operated at CDS, Strasbourg, France.
% for native speakers
The authors received editing help from Travis Thieme, a Ph.D. student in National Tsing Hua University, who has English fluency and knowledge in astrophysics.

%%%%%%%%%%%%%%%%%%%%%%%%%%%%%%

%%%%%%%%%%%%%%%%%%%%%%%%%%%%%%%%%%%%%%%%%%%%%%%%%%%
%%%%%%%%%%%%%%%%%%%%%%%%%%%%%%%%%%%%%%%%%%%%%%%%%%%
\appendix

\section{Hyper-Parameters and Comparison of Machine Learning Algorithms}
\label{Appendix: Comparison_ML}

%========================
\begin{figure}[htb!]
\centering
\includegraphics[width=0.96\textwidth]{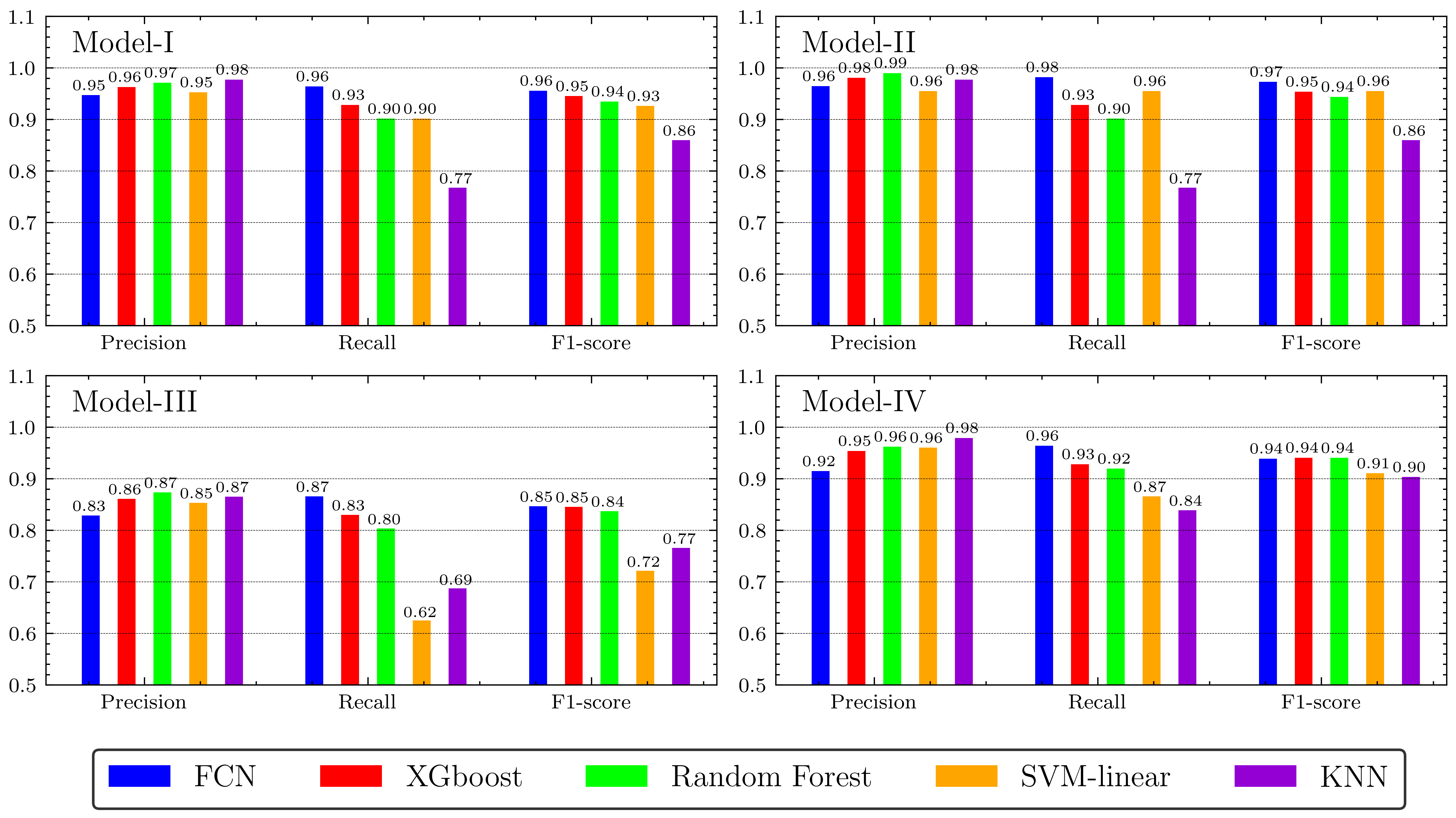}
\caption{The recall, precision and F1-scores of five different machine learning algorithms in the four models described in the manuscript. Model I uses the full(8) SED bands as the input features, Model II uses full SED bands as well as their errors, Model III uses normalized SED bands, and Model IV uses the three bands in the long wavelength region. Hyper-parameters used in these algorithms are described in Table \ref{tab:ML_setting}}

\label{fig:Prediction_total.png}
\end{figure}
%=======================

In this appendix, we provide the comparisons of five machine learning algorithms in the four models (Model I-IV) described in the manuscript. These five algorithms are Fully Connected Neural Network (FCN), eXtreme Gradient Boosting (XGB), Random Forest (RF), Support Vector Machine (SVM), and k-nearest neighbor (KNN).
As shown in Figure \ref{fig:Prediction_total.png}, they have similar results in most of the results. However, we do find that FCN is slightly better than others, because it provides the highest F1-score and recall in all of the four models. XGB and RF have similar results to the FCN and thus could be the potential candidates too, while SVM have worse performance than the former in Model III. This implies that the classification of SVM is mainly based on the magnitude (i.e. distance effects) rather than the SED pattern. On the other hand, KNN are apparently worse than all others, except for in the model IV. It may imply again that the sources could potentially be classified by clustering through the last three bands in the long wavelength region (see Figure \ref{fig:SED 3D}).

The implementation of these machine learning algorithms are scikit-learn and tensorflow, which are the common packages in Python. Table \ref{tab:ML_setting} lists the hyper-parameters we used for these models. For XGB, RF and KNN, we choose appropriate parameters (after more comparisons with other possible combinations, not shown here) and make sure that the performance cannot be improved even for more parameters. As for SVM, only results of linear kernel can have the precise prediction and are therefore shown here. Other kernels (sigmoid, polynomial and radial basis function) 
are all inappropriate, because the obtained recall of YSO are even lower than $20\%$. 

%======================
\begin{table}[h]
\centering
\begin{tabular}{c|c}
\hline\hline
Model & Hyper-parameters \\
\hline
XGB & n\_estimators=100, max\_depth=15 \\
RF & n\_estimators=100, max\_depth=15 \\
SVM & kernel='linear', gamma='auto' \\
KNN & n\_neighbors=6 \\
\hline
\hline
\end{tabular}
\caption{The hyper-parameters of machine learning models. Others we do not mentioned are the defaults for scikit-learn package.}
\label{tab:ML_setting}
\end{table}
%================

%%%%%%%%%%%%%%%%%%%%%%%%%%%%%%%%%%%%%%%%%%%%%%%%%%%%%%%%%
\section{SCAO prediction list from \textit{Spitzer} enhanced imaging products}
\label{Appendix: MRT for SEIP}
%===========
\begin{table}[h]
\centering
\tiny
\begin{tabular}{cc|cccccccc|cccccccc}
\hline\hline
\multicolumn{2}{c|}{Coordinate}&\multicolumn{8}{c|}{Model II}&\multicolumn{8}{c}{Model IV}\\
\hline
R.A. & Dec. & Star & Galaxy & YSOc & FCN & XGB & RF & SVM & KNN & Star & Galaxy & YSOc & FCN & XGB & RF & SVM & KNN\\
(deg.) & (deg.) & prob & prob & prob & pred & pred & pred & pred & pred & prob & prob & prob & pred & pred & pred & pred & pred \\
\hline 
1.198290 & 68.515790 & 0.000 & 0.277 & 0.723 & Y & Y & S & Y & S & 0.002 & 0.000 & 0.998 & Y & Y & Y & S & S \\ 
3.055216 & -73.027300 & &&&&&&&& 0.044 & 0.956 & 0.000 & G & G & S & S & G \\
\hline
\hline
\end{tabular}
\caption{Examples of SCAO prediction list. Each row represents a source in SEIP catalog, and columns respectively show the coordinates in the SEIP catalog, probabilities obtained from Model II, and probabilities obtained from Model IV. 
If the label of a source can only be predicted by one model, the other probability column is left blank.}
\tablecomments{This table is available in its entirety in a machine-readable form in the online journal. A portion is shown here for guidance regarding its form and content.}
\label{tab:MRT}
\end{table}
%========

We apply SCAO to the SEIP, which is the most complete \textit{Spitzer} source catalog, and provide a catalog list in 
Table \ref{tab:MRT}.
We only use Model II and Model IV since Model II makes the most accurate predictions and Model IV provides the largest number of predictions among our models. For each source, we provide the average probabilities from ten runs of FCN and the predictions from five different machine learning models. The labels S,G and Y represent star, galaxy and YSO candidates, respectively.

%%%%%%%%%%%%%%%%%%%%%%%%%%%%%%%%%%%%

\begin{thebibliography}{}

\bibitem[Akras et al.(2019)]{Akras2019} Akras, S., Leal-Ferreira, M.~L., Guzman-Ramirez, L., et al.\ 2019, \mnras, 483, 5077

\bibitem[Kingma \& Ba(2015)]{Adam}
Kingma D. P., \& Ba J. \ 2015, 3rd International Conference for Learning Representations, San Diego (arXiv:1412.6980)

\bibitem[An et al.(2011)]{An2011} An, D., Ram{\'\i}rez, S.~V., Sellgren, K., et al.\ 2011, \apj, 736, 133

\bibitem[Bergin \& Tafalla(2007)]{Bergin2007} Bergin, E.~A., \& Tafalla, M.\ 2007, \araa, 45, 339.

\bibitem[Casali et al.(2007)]{Casali2007} Casali, M., Adamson, A., Alves de Oliveira, C., et al.\ 2007, \aap, 467, 777

\bibitem[Chiu \& Lai(in prep.)]{Chiu_in_prep} Chiu, Y.-L., \& Lai, S.-P. \ 2020, in preparation

\bibitem[de Villiers(2013)]{deVilliers2013} de Villiers, H.\ 2013, Protostars and Planets VI Posters

\bibitem[Draine(2003)]{Draine2003} Draine, B.~T.\ 2003, \araa, 41, 241

\bibitem[Evans et al.(2003)]{Evans2003} Evans, N.~J., II, Allen, L.~E., Blake, G.~A., et al.\ 2003, \pasp, 115, 965

\bibitem[Evans et al.(2007)]{Evans2007} Evans, N. J., II, Harvey, P. M., Dunham, M. M., et al. 2007, Final Delivery of Data from the c2d Legacy Project: IRAC and MIPS (Pasadena, CA: SSC)

\bibitem[Evans et al.(2009)]{Evans2009} Evans, N.~J., II, Dunham, M.~M., J{\o}rgensen, J.~K., et al.\ 2009, \apjs, 181, 321

\bibitem[Furlan et al.(2016)]{Furlan2016} Furlan, E., Fischer, W.~J., Ali, B., et al.\ 2016, \apjs, 224, 5

\bibitem[Greene et al.(1994)]{Greene1994} Greene, T.~P., Wilking, B.~A., Andre, P., et al.\ 1994, \apj, 434, 614

\bibitem[Hambly et al.(2008)]{Hambly2008} Hambly, N.~C., Collins, R.~S., Cross, N.~J.~G., et al.\ 2008, \mnras, 384, 637.

\bibitem[Harvey et al.(2007)]{Harvey2007} Harvey, P., Mer{\'\i}n, B., Huard, T.~L., et al.\ 2007, \apj, 663, 1149

\bibitem[Hedges et al.(2018)]{Hedges2018} Hedges, C., Hodgkin, S., \& Kennedy, G.\ 2018, \mnras, 476, 2968

\bibitem[Hewett et al.(2006)]{Hewett2006} Hewett, P.~C., Warren, S.~J., Leggett, S.~K., \& Hodgkin, S.~T.\ 2006, \mnras, 367, 454 

\bibitem[Hodgkin et al.(2009)]{Hodgkin2009} Hodgkin, S.~T., Irwin, M.~J., Hewett, P.~C., \& Warren, S.~J.\ 2009, \mnras, 394, 675

\bibitem[Hsieh \& Lai(2013)]{Hsieh2013} Hsieh, T.-H., \& Lai, S.-P.\ 2013, \apjs, 205, 5

\bibitem[Irwin et al.(in prep.)]{Irwin_in_prep} Irwin et al., 2008, in preparation

\bibitem[Koenig \& Leisawitz(2014)]{Koenig2014} Koenig, X.~P., \& Leisawitz, D.~T.\ 2014, \apj, 791, 131

\bibitem[LeCun et al. (2015)]{ANN}
LeCun, Y., Bengio, Y., and Hinton, G. \ 2015, Nature 521, 436–444

\bibitem[Lombardi \& Alves(2001)]{Lombardi2001} Lombardi, M., \& Alves, J.\ 2001, \aap, 377, 1023.

\bibitem[Lonsdale et al.(2003)]{Lonsdale2003} Lonsdale, C.~J., Smith, H.~E., Rowan-Robinson, M., et al.\ 2003, \pasp, 115, 897 

\bibitem[Gutermuth et al.(2005)]{Gutermuth2005} Gutermuth, R.~A., Megeath, S.~T., Pipher, J.~L., et al.\ 2005, \apj, 632, 397

\bibitem[Lawrence et al.(2007)]{Lawrence2007} Lawrence, A., Warren, S.~J., Almaini, O., et al.\ 2007, \mnras, 379, 1599

\bibitem[Marton et al.(2016)]{Marton2016} Marton, G., T{\'o}th, L.~V., Paladini, R., et al.\ 2016, \mnras, 458, 3479

\bibitem[Marton et al.(2019)]{Marton2019} Marton, G., {\'A}brah{\'a}m, P., Szegedi-Elek, E., et al.\ 2019, \mnras, 487, 2522

\bibitem[McCulloch \& Walter(1943)]{ANN_early}
McCulloch, W. \& Walter P. \ 1943, Bulletin of Mathematical Biophysics. 5 (4): 115–133

\bibitem[McKee \& Ostriker(2007)]{McKee2007} McKee, C.~F., \& Ostriker, E.~C.\ 2007, \araa, 45, 565.

\bibitem[Miettinen(2018)]{Miettinen2018} Miettinen, O.\ 2018, \apss, 363, 197

\bibitem[Meingast et al.(2017)]{Meingast2017} Meingast, S., Lombardi, M., \& Alves, J.\ 2017, \aap, 601, A137

\bibitem[Rebull et al.(2010)]{Rebull2010} Rebull, L.~M., Padgett, D.~L., McCabe, C.-E., et al.\ 2010, \apjs, 186, 259 

\bibitem[Russell et al. (2014)]{Russell2014}
See for example, Artificial Intelligence: A Modern Approach, by Stuart Russell and Peter Norvig (Third Edition, Pearson Education 2014).

\bibitem[Skrutskie et al.(2006)]{Skrutskie2006} Skrutskie, M.~F., Cutri, R.~M., Stiening, R., et al.\ 2006, \aj, 131, 1163

\bibitem[Wenger et al.(2000)]{Wenger2000} Wenger, M., Ochsenbein, F., Egret, D., et al.\ 2000, Astronomy and Astrophysics Supplement Series, 143, 9.

\bibitem[Werner et al.(2004)]{Werner2004} Werner, M.~W., Roellig, T.~L., Low, F.~J., et al.\ 2004, \apjs, 154, 1 

\end{thebibliography}
\end{document}